\documentclass[twocolumn,aps,prl,groupedaddress]{revtex4-1}
\usepackage{amsmath,amssymb}
\usepackage{graphicx,xcolor,soul,sidecap}
\usepackage[colorlinks,bookmarks=false,citecolor=blue,linkcolor=blue,urlcolor=blue,hyperfootnotes=true]{hyperref}
\makeatother

\begin{document}
\title{Tunable critical field in Rashba superconductor thin-films}

\author{L. A. B. Olde Olthof} 
\author{J. R. Weggemans} 
\author{G. Kimbell} 
\author{J. W. A. Robinson} \email[]{jjr33@cam.ac.uk}
\author{X. Montiel} \email[]{xm252@cam.ac.uk}
\affiliation{Department of Materials Science \& Metallurgy, University of Cambridge, CB3 0FS Cambridge, United Kingdom}
\date{\today}
\begin{abstract}
The upper critical field in type~II superconductors is limited by the Pauli paramagnetic limit. 
In superconductors with strong Rashba spin-orbit coupling this limit can be overcome by forming a helical state.
Here we quantitatively study the magnetic field-temperature phase diagram of finite-size superconductors with Rashba spin-orbit coupling.
We discuss the effect of finite size and shape anisotropy.
We demonstrate that the critical field is controllable by intrinsic parameters such as spin-orbit coupling strength and tunable parameters such as sample geometry and applied field direction. Our study opens new avenues for the design of superconducting spin-valves.
\end{abstract}
\maketitle

In spin-singlet superconductors, an applied magnetic field exceeding the upper critical field~$h_{c2}$ destroys superconductivity by means of orbital \cite{Degennes} and Pauli paramagnetic effects \cite{Clogston,Chandra}.
In a thin-film geometry, the orbital contribution is negligible \cite{Tinkham}. A singlet Cooper pair breaks when the binding energy is exceeded by the Zeeman splitting energy, as defined by the Clogston-Chandraskhar or Pauli paramagnetic limit \cite{Clogston,Chandra}. The corresponding critical field~$h_p$ is the zero-temperature limit of $h_{c2}$. 
In the presence of only the paramagnetic effect, $h_{c2}$ is a first-order transition at low temperature and becomes second-order at higher temperature \cite{Sarma,Maki64,MakiTsuneto}.

During the past decades, several methods have been explored to overcome $h_p$.
Spin-orbit scattering randomizes spins that scatter off boundaries \cite{Ferrel} or impurities \cite{Anderson,AbrikosovGorkov,Klemm,Grimaldi}, which lowers the spin susceptibility and weakens the paramagnetic effect.
In the Larkin-Ovchinnikov-Fulde-Ferrell (LOFF) state, Cooper pairs acquire a finite momentum and the pair wave function is spatially modulated \cite{LO,FF,Matsuda}. 
The LOFF state is stable in the clean limit and disappears in the presence of impurities
\cite{Aslamazov_JETP1969,Bulaevskii_1976}.
In spin-triplet superconductors, spin-aligned Cooper pairs are unaffected by paramagnetism \cite{Aoki2001,Huy,Aoki2012}.

\begin{figure}
    \centering
    \includegraphics[width=.75\linewidth]{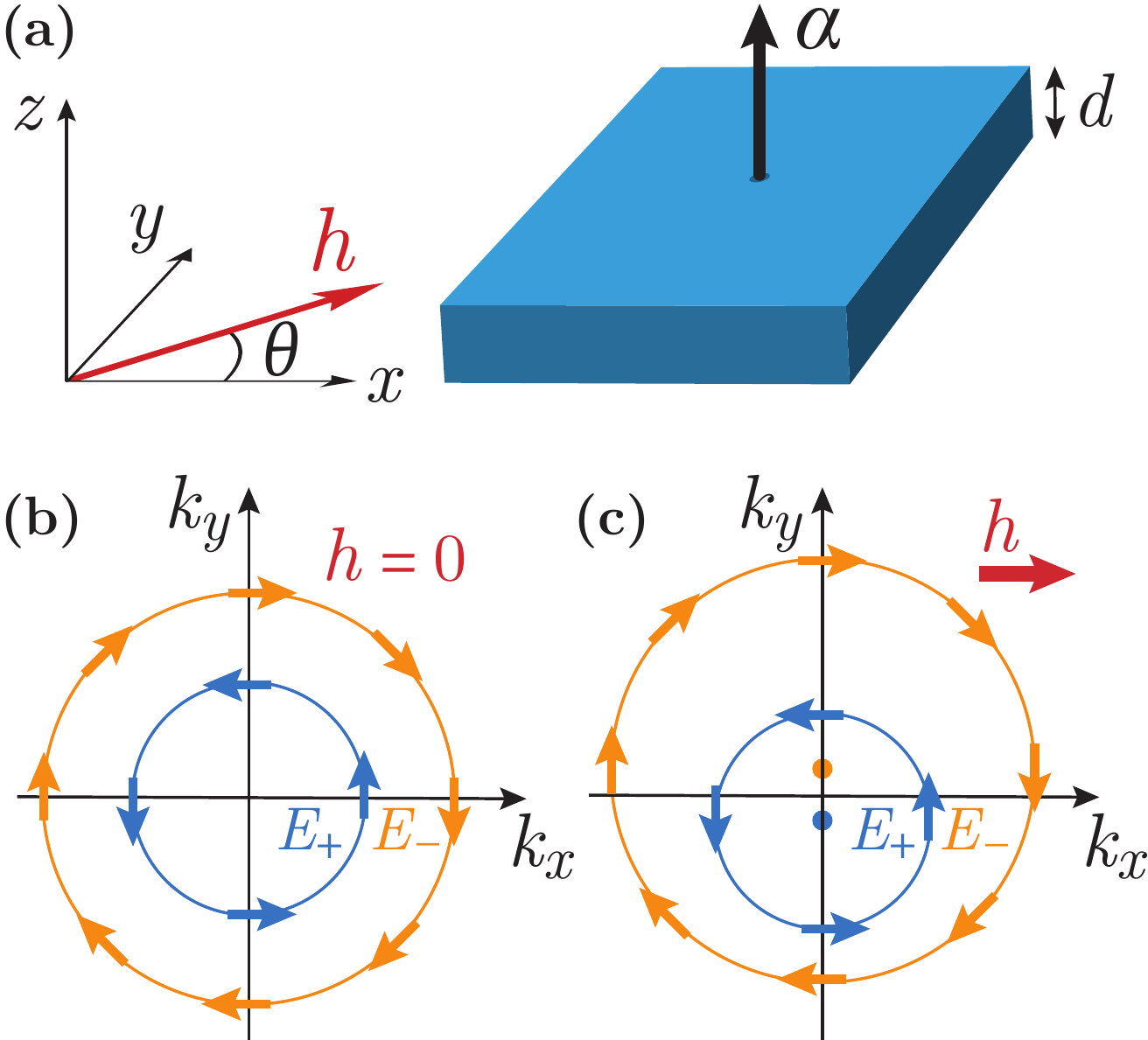}
    \caption{{\bf (a)} Schematic illustration of a thin-film superconductor with thickness $d$, out-of-plane spin-orbit coupling $\vec{\alpha}$ in an applied in-plane magnetic field $\vec{h}$.
    {\bf (b)} Fermi surface of a Rashba superconductor. The spin is locked to the momentum, forming two helicity bands $E_+$ and $E_-$. 
    {\bf (c)} An applied magnetic field $\vec{h}=(h_x,0,0)$ shifts the helicity bands vertically, the dots representing their new centers. In the presence of a field, the Rashba superconductor has intrinsic spatial anisotropy.}
    \label{fig:Rashba}
\end{figure}

Here, we focus on thin-film superconductors with Rashba spin-orbit coupling (SOC).
The usual quadratic dispersion is split into two helicity bands with energies $E_\pm = \hbar k^2/2m \pm \alpha |k|$, where $\alpha$ is the SOC strength and $k$ the single-particle momentum \cite{Smidman}. The spins are polarized tangential to their momentum, known as spin-momentum locking, as illustrated in Fig.~\ref{fig:Rashba}(b). For each direction in momentum space, there are two zero-momentum opposite-spin Cooper pairs on the Fermi surface.
Adding an in-plane magnetic field $\vec{h}=(h_x,h_y,0)$, the dispersion becomes $E_\pm = \hbar k^2/2m \pm \sqrt{ (\alpha k_y + h_x)^2 + (\alpha k_x-h_y)^2}$ \cite{Smidman}. The Fermi surfaces shift in the direction perpendicular to the field, producing an intrinsic spatial anisotropy  [see Fig~\ref{fig:Rashba}(c)]. Consequently, the singlet Cooper pairs acquire a net momentum and a LOFF-like state forms in the clean limit \cite{Barzykin,Agterberg2007,Loder}. The critical field experiences a sharp incline at low temperatures, surpassing the paramagnetic limit \cite{Agterberg2007,Dimitrova}.
In the diffusive regime, the LOFF-like state disappears and a spatially modulated helical state remains \cite{Dimitrova,Houzet}. This state is stable against disorder since it originates from the SOC symmetry \cite{Agterberg2003}.
Disordered Rasbha superconductors with strong SOC in an in-plane magnetic field thus have an enhanced critical field \cite{Dimitrova,Houzet} and critical temperature \cite{Gardner}.

In this {\it Letter}, we study the effect of the SOC-induced anisotropy (as illustrated in Fig.~\ref{fig:Rashba}) on the magnetic field-temperature $(h,T)$ phase diagram of finite sized Rashba superconductors. 
In particular, we theoretically show: i) a significant enhancement of the paramagnetic limit of singlet superconductors in the presence of Rashba SOC; ii) in Rashba superconductors, the paramagnetic limit strongly depends on the geometry of the sample; and iii) the paramagnetic limit is controlled by the orientation of the applied magnetic field in geometrically anisotropic thin-film superconductors.

We investigate superconducting thin-films with in-plane magnetic field $\vec{h}$ (externally applied or via an induced exchange field). We model superconductivity in the diffusive limit via the Usadel formalism, which is formulated in terms of Green's functions \cite{Champel}. 
The Green's functions $\hat{g}(\vec{R},\omega_n)$ depend on the spatial coordinate $\vec{R}$ and the Matsubara frequencies $\omega_n = (2n+1)\pi T$ ($T$ is the temperature and $n\in\mathbb{Z}$);
in $4\otimes 4$ spin $\otimes$ particle-hole space, $\hat{g}(\vec{R},\omega_n)$ is expressed as \cite{Champel}
\begin{equation}
\hat{g}=\left(\begin{array}{cc}
g & f\\
\widetilde{f} & \widetilde{g}
\end{array}\right) ,
\end{equation}
where $g$ and $f$ are the normal and anomalous Green's functions, respectively, and $\widetilde{g}$ and $\widetilde{f}$ are their particle-hole conjugates \cite{Champel}. In the diffusive limit, the Green's functions satisfy the Usadel transport equation \cite{Champel}
\begin{equation}
    \left[ i\omega_n\hat{\tau}_z -\hat{\Delta}-\vec{h}\cdot\vec{\sigma},\hat{g}\right] + \frac{D}{\pi}\nabla(\hat{g}\nabla\hat{g})=0,
    \label{eq_usadel}
\end{equation}
with the normalization condition $\hat{g}^2=-\pi^2\hat{1}$.
In Eq.~(\ref{eq_usadel}), $\vec{\sigma}$ and $\vec{\tau}$ are the Pauli matrices in spin and particle-hole space, respectively, $D$ is the diffusion coefficient and $\hat{\Delta}= \Delta_s i\sigma^y$ is the conventional superconducting order parameter.
The Rashba SOC gives rise to an effective momentum-dependent exchange field, i.e. the spin-orbit field. To include this in the Usadel equations, we introduce the covariant derivative
$\bar{\nabla} \mapsto \nabla - i [ \underline{\hat{A}}, .]$,
where $\nabla$ is the standard derivative and  $\underline{\hat{A}}$ the spin-orbit field vector \cite{Bergeret,Jacobsen,Montiel}. In the following, we assume that the spin-charge conversion terms are negligible
\footnote{A quantitative study of spin-charge conversion in Rashba superconductors is given in Ref.~\cite{Bergeret2020}.}. 
Close to the critical temperature $T_\text{c}$, the Usadel equations become \cite{Champel}
\begin{equation}
    \begin{cases}
    \left( D\bar{\nabla}^2 - 2|\omega_n|\right)f_s = - 2\pi\Delta_s + 2i\mbox{ sgn}(\omega_n) \vec{h}\cdot \vec{f}_t, \\
    \left( D\bar{\nabla}^2 - 2|\omega_n|\right)\vec{f}_t = 2i\mbox{ sgn}(\omega_n) \vec{h} f_s,
    \end{cases} \label{UsadelTc}
\end{equation}
where the normal Green's function is $g=-i\pi \tau_3$ and the anomalous Green's function is decomposed in the spin Pauli matrices base as $f=(f_s+\vec{f}_t\cdot\vec{\sigma})i\sigma^y$ where $f_s$ is the singlet pair amplitude and $\vec{f}_t = (f_t^x,\ f_t^t,\ f_t^z)^T$ the triplet one.

In the following, we study superconducting thin-films lying in the $xy$-plane with thickness $d$ 
smaller than the superconducting coherence length, i.e. $d\leq \xi$, with $\xi=\sqrt{D/2\pi T_c}$. We assume that the superconductivity is uniform in~$z$, such that the Green's functions only depend on the $x$ and $y$  coordinates, i.e. $\hat{g}(\vec{R},\omega_n)=\hat{g}(x,y,\omega_n)$.
The spin-orbit field is $\vec{\alpha} = \alpha (\vec{s}\times\vec{p})\cdot\hat{n}$, where $\alpha$ is the SOC strength (in units of 1/$\xi$), the spin $\vec{s} = |\vec{h}|(\cos\theta,\ \sin\theta,\ 0)^T$ is determined by the in-plane field $\vec{h}$ (see Fig.~\ref{fig:Rashba}(a)), the momentum in a thin-film is $\vec{p} = (p_x,\ p_y,\ 0)^T$ and the unit vector along the axis of broken symmetry is $\hat{n}=\hat{z}$.
Hence, the spin-orbit field becomes $\vec{\alpha} = \alpha\left( h_xp_y - h_yp_x \right)\hat{z}$ and is directed out-of-plane.
The corresponding spin-orbit field coefficients in spin-space are $A_x = - \alpha \sigma^y$, $A_y = \alpha\sigma^x$ and $A_z=0$, which are used in the covariant derivative $\bar{\nabla}$ (see \cite{supp} for details).

\begin{figure}
    \centering
    \includegraphics[width=\linewidth]{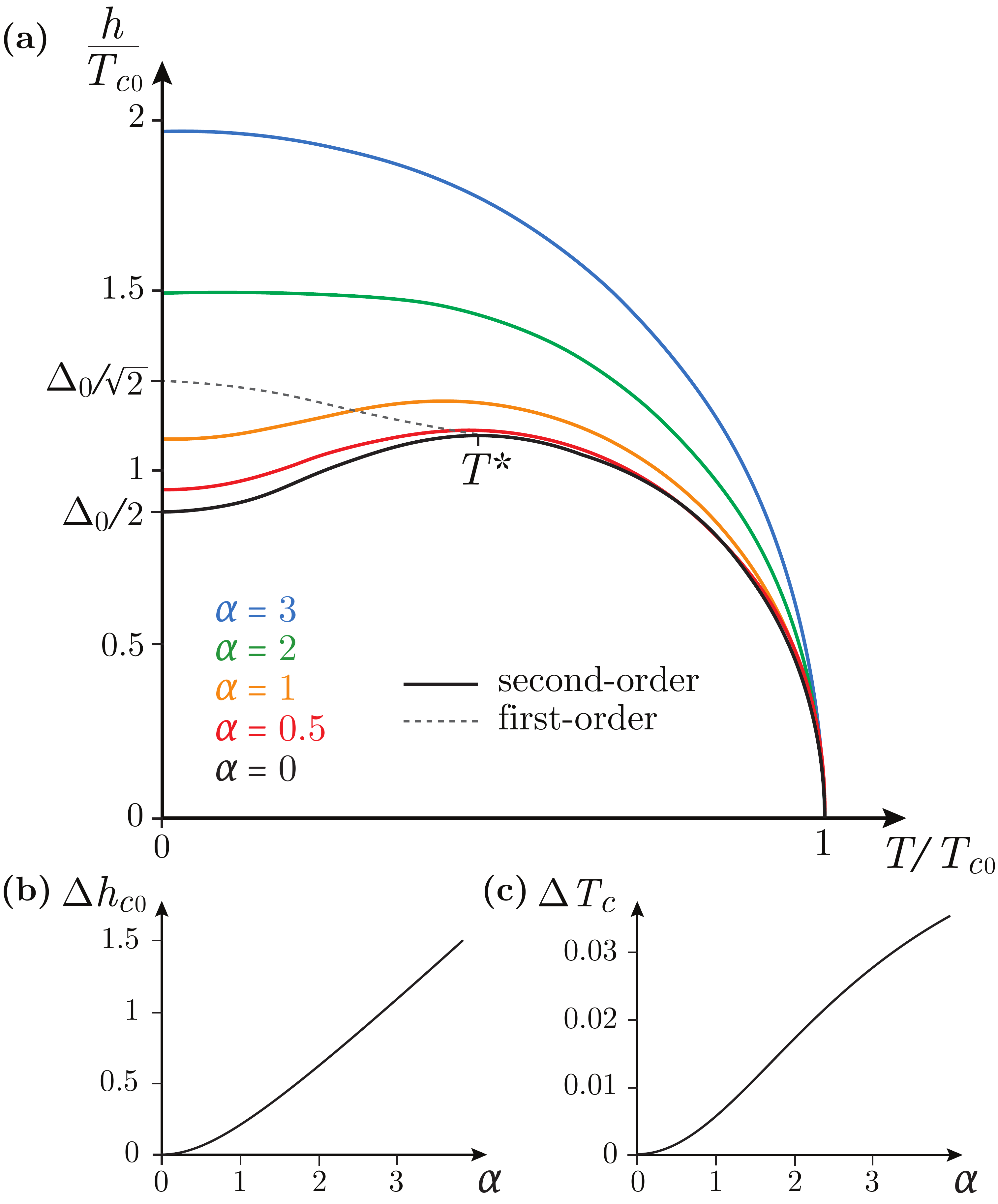}
    \caption{Properties of an infinite thin-film Rashba superconductor.
    {\bf (a)} Phase diagram for different values of Rashba spin-orbit coupling strength $\alpha$. Solid lines are second-order self-consistent transitions, meaning that the order parameter vanishes at  $\Delta(T=T_\text{c})=0$.
    The dashed line is the first-order paramagnetic limit at $\alpha=0$. Both phase transition lines meet at the tricritical point at $T=T^*$ \cite{SaintJames}. For $T<T^*$, the second order phase transition defines the supercooling magnetic field. 
    {\bf (b)} The increase in critical field at zero temperature $\Delta h_{c0} = |h_{c0}(\alpha) - h_{c0}(\alpha=0)|/T_{c0}$ with $\alpha$.
    {\bf (c)} The increase in critical temperature $\Delta T_\text{c} = |T_\text{c}(\alpha) - T_\text{c}(\alpha=0)|/T_{c0}$ with $\alpha$, for fixed applied field $h/T_{c0}=0.5$.}
    \label{fig:phase1}
\end{figure}

We first investigate the $(h,T)$ phase diagram for an infinite in $(x,y)$ thin-film Rashba superconductors. In infinite films, the spatial derivative in Eq.~(\ref{UsadelTc}) can be neglected. We derive the self-consistency equation and solve it analytically \cite{supp} to map the phase diagram in Fig.~\ref{fig:phase1}(a).
With increasing SOC strength $\alpha$, the critical field increases and the transition takes on a concave shape, similar to~\cite{Dimitrova}.
The spin-momentum locking caused by SOC renormalizes the magnetic field, meaning that with increasing $\alpha$, the effective $\vec{h}$ decreases \cite{supp}.
At zero magnetic field, SOC does not affect $T_\text{c}$, showing that SOC does not affect the superconductivity, but screens the applied magnetic field \cite{note_SOC}. A similar screening effect is observed in the presence of spin-orbit scattering in disordered superconductors \cite{AbrikosovGorkov,Grimaldi}. 

SOC increases the critical field, as shown at zero temperature in Fig.~\ref{fig:phase1}(b). 
The magnitude of $\Delta T_\text{c}$ at finite magnetic field, seen in Fig.~\ref{fig:phase1}(c), is similar to the temperature recovery predicted in superconductor/ferromagnet bilayers \cite{Olde}. 
We note that the largest change in $h_{c2}$ is two orders of magnitude higher than the change in $T_c$, implying that the effect of SOC on magnetic field is more easily observable.

\begin{figure}
    \centering
    \includegraphics[width=\linewidth]{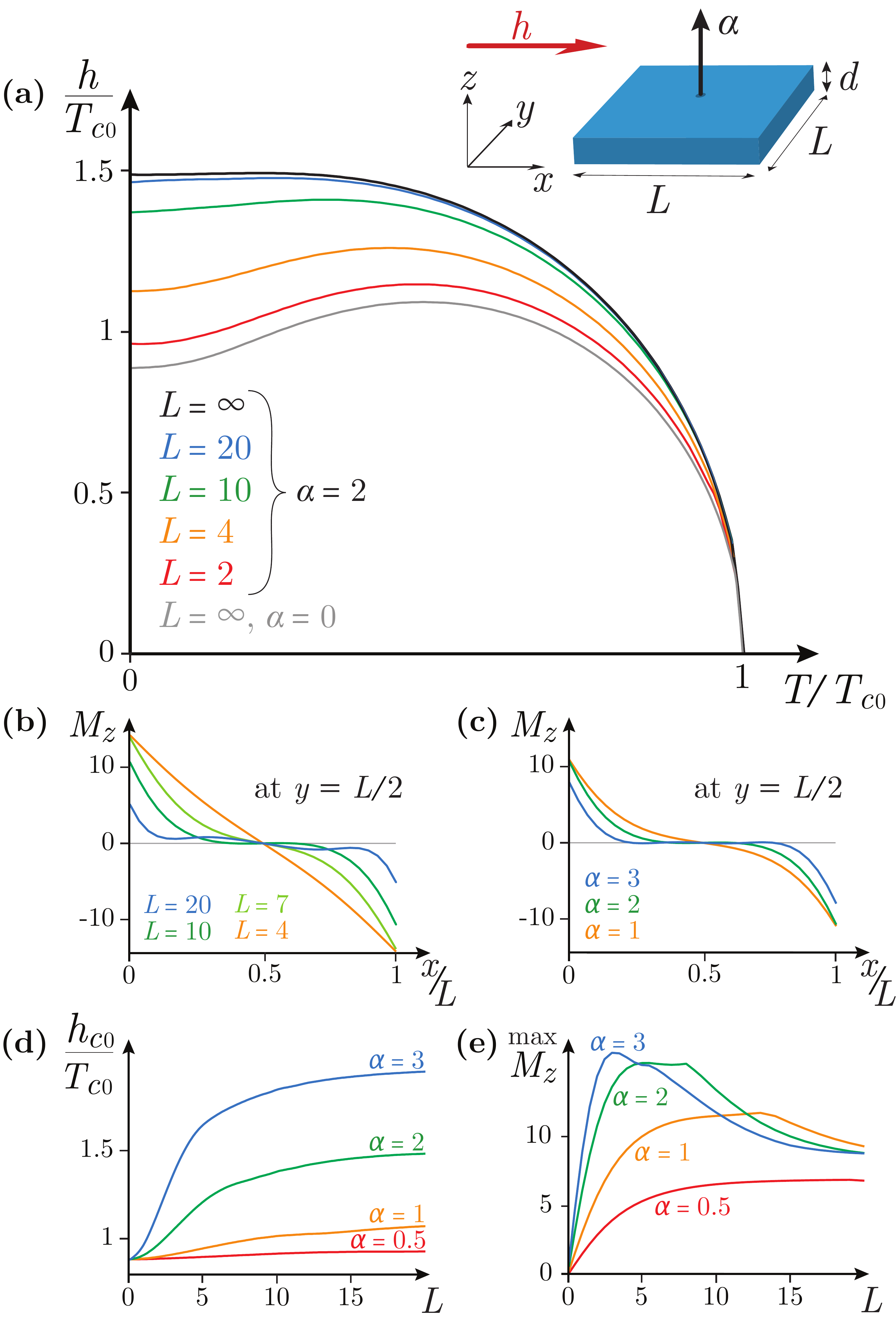}
    \caption{The effect of finite size.
    Top-right inset: Geometrically constrained $L\times L$ thin-film superconductor with out-of-plane spin-orbit coupling $\vec{\alpha}$ in an applied magnetic field $\vec{h}=(h_x,0,0)$.
    {\bf (a)} Phase diagram of a $L\times L$ superconductor with $\alpha=2$ (colored), along with the analytical infinite film solutions for $\alpha = 2$ (black) and $\alpha = 0$ (grey).
    {\bf (b)} The profile of the induced out-of-plane spin magnetization $M_z$ in the field direction (along~$x$) in the middle of the sample ($y=L/2$), for fixed $\alpha=2$ and different values of $L$.
    {\bf (c)} The same profile for fixed $L=10$ and different values of $\alpha$.
    {\bf (d)} The critical field at zero temperature $h_{c0}$ and
    {\bf (e)} the maximum of $M_z$ as a function of $L$, for different values of $\alpha$.}
    \label{fig:phase2}
\end{figure}

While the SOC screens the magnetic field in infinite thin-films, an additional effect appears at the edge of finite samples. Edge states with distinct physical properties than the infinite film superconductors may appear as in topological superconductors \cite{Menard,Bergeret2020}.
We consider a finite $L_x\times L_y$ superconductor, where $L_x$ and $L_y$ are in units of $\xi$. We introduce the boundary condition  that the transverse current vanishes at the edge of the sample, implying that the covariant derivatives vanish \cite{supp}.
The phase diagram is calculated iteratively, starting from the analytical infinite film solution as an Ansatz \cite{supp}.

The numerical phase diagram for a $L\times L$ superconductor is shown in Fig.~\ref{fig:phase2}(a). The length $L$ is in units of~$\xi$, such that $L=20$ converges to the analytical infinite film solution. Decreasing $L$ reduces $h_{c2}$ compared to the corresponding infinite film value.

Microscopically, the in-plane field $\vec{h}$ fixes the spin quantization axis along the field direction $\hat{h}$, which causes spin mixing and forms short-range zero spin triplets, described by $f_\text{SR} = \vec{f}_t\cdot\hat{h}$ \cite{Jacobsen}. Adding Rashba SOC rotates the spins out-of-plane, resulting in long-range equal-spin triplets, described by the projection $f_\text{LR} = |\vec{f}_t\times\hat{h}|$ perpendicular to the field \cite{Jacobsen}.
Hence, the joint action of the SOC and $\vec{h}$ in finite-size samples results in the spins having a preferred alignment, i.e. an induced spin magnetization in the superconductor \cite{Bergeret2020}, given by
\begin{equation}
    \vec{M}(x,y) = (M_x,M_y,M_z) = M_0 \sum_n f_s \vec{f}_t,
\end{equation}
where $M_0$ is a constant defined in~\cite{supp} and the summation is over the Matsubara frequencies.

The out-of-plane spin magnetization $M_z$ is introduced via the boundary conditions as an edge effect with a characteristic length scale $\xi$, meaning that it dominates when $L\sim\xi$ and does not exist in infinite films \cite{Bergeret2020}.
The vanishing of covariant derivative at the edges directly couples the out-of-plane triplet pair amplitude with short-range triplet correlations \cite{supp}.

An applied field $\vec{h}=(h_x,0,0)$ acting on spins pointing along $\pm k_y$ generates a magnetization along $\pm z$. 
Due to spin-momentum locking [illustrated in Fig.~\ref{fig:Rashba}(b)] a net magnetization occurs in opposite directions on either side of the Fermi surface.
The resulting magnetization profile is positive on one side of the sample, zero in the middle and negative on the other side~\cite{Bergeret2020}, as shown in Fig.~\ref{fig:phase2}(b)-\ref{fig:phase2}(c).
The magnetization acquires this profile in the field direction (along~$x$), whilst remaining nearly constant in the perpendicular direction (along~$y$).
For small $L$, a magnetization gradient spans the whole sample. Upon increasing $L$, the magnetization becomes concentrated at the edges. A similar effect is seen when increasing $\alpha$.
The profile resembles that of the spin-orbit induced local magnetic field in a superconductor/ferromagnet bilayer and could therefore lead to the formation of vortices \cite{Olde}.

When the system becomes small ($L\sim\xi$), the boundary conditions at the the edges dominate the sample and the covariant derivative vanishes everywhere, resulting in a recovery of the infinite film phase diagram in the absence of SOC \cite{supp} [see Fig.~\ref{fig:phase2}(a)].
We thus conclude that the SOC gives rise to two competing effects: the infinite film screening effect (increasing $h_{c2}$) and the edge effect (suppressing $h_{c2}$).

The critical field $h_{c0}$ and magnetization $M_z$ are shown as a function of $L$ for different values of $\alpha$ in Fig.~\ref{fig:phase2}(d)-\ref{fig:phase2}(e).
In small samples ($L\sim\xi$), the edge effect dominates. As a result, $h_{c0}$ and $M_z$ rapidly increase with~$L$.
For large $L$, $h_{c0}$ saturates and $M_z$ gradually drops off to a residual magnetization which is not present in infinite films (in which $M_z=0$).

\begin{figure}
    \centering
    \includegraphics[width=\linewidth]{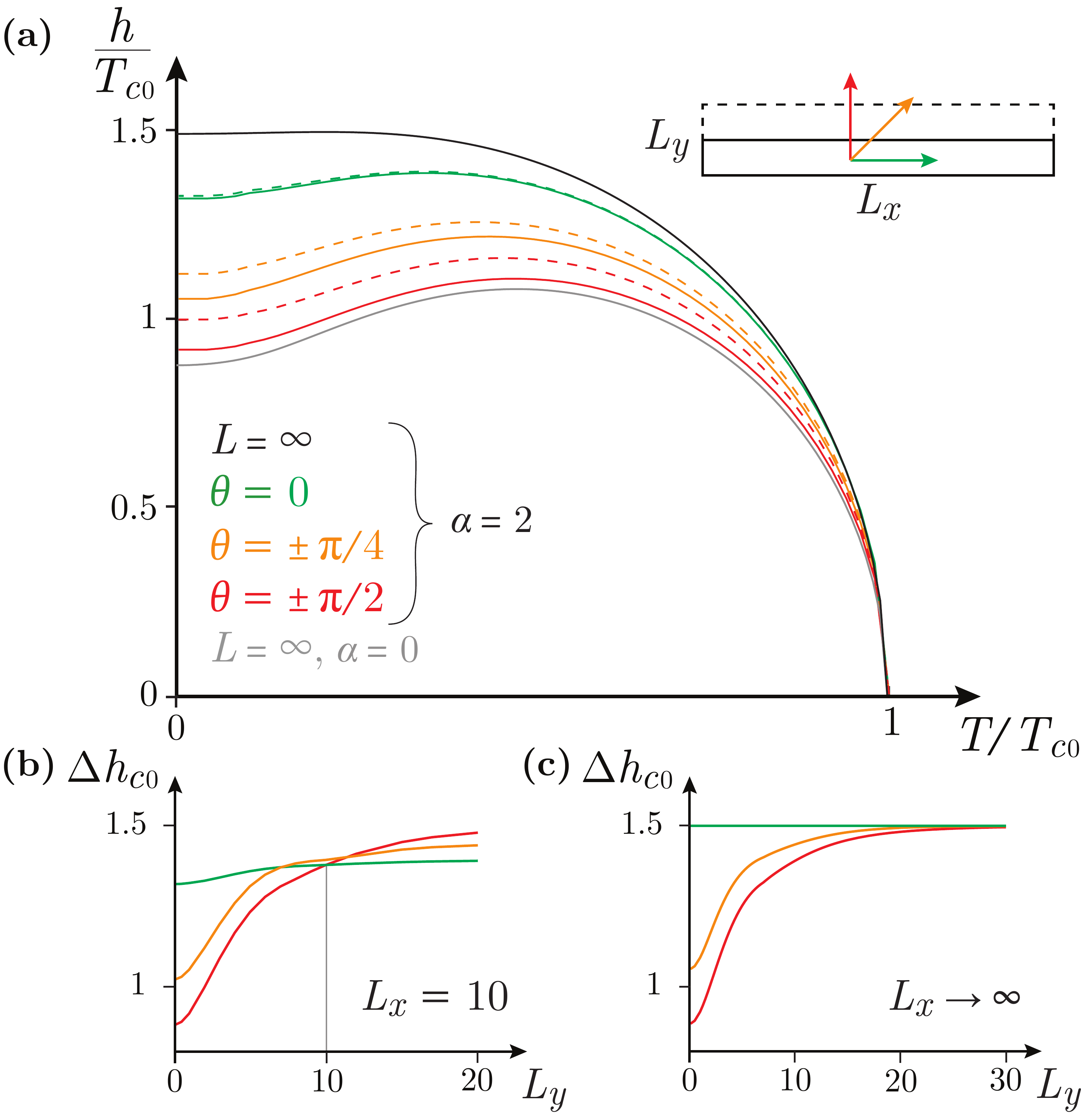}
    \caption{The effect of shape anisotropy. 
    {\bf (a)} Phase diagram of a rectangular superconductor with $\alpha =2$, $L_x = 10 L_y$ (solid) and $L_x = 5L_y$ (dashed), along with infinite film solutions for $\alpha = 2$ (black) and $\alpha = 0$ (grey). Rotating the in-plane magnetic field angle $\theta$ along the long axis of the sample increases the critical field. 
    {\bf (b)} The effect of anisotropy on the critical field $h_{c0}$.
    {\bf (c)} The one-dimensional limit.}
    \label{fig:phase3}
\end{figure}

To investigate further the edge effect, we calculate the phase diagram of a rectangular superconductor with $L_x>L_y$.
The shape anisotropy introduces an in-plane angle $\theta$ between $\vec{h}$ and the $x$-axis.
When $\vec{h}$ points along the larger dimension ($x$), the edge magnetization is strong along the shorter dimension ($y$), and vice versa. 
Therefore, the balance between the edge effect and the magnetic field screening effect is controllable by sample geometry in combination with the applied field direction, as shown in Fig.~\ref{fig:phase3}(a). When the magnetic field points along the larger dimension ($\theta=0$), $h_{c2}$ is only slightly suppressed compared to the infinite film with the same value of $\alpha$. 
Upon rotating $\theta$, $h_{c2}$ gets suppressed further and has a minimum for $\theta=\pm\pi/2$, where it approaches the infinite film in the absence of SOC.

The quantitative effect of the field direction on a superconductor with constant $L_x$ and increasing $L_y$ is shown in Fig.~\ref{fig:phase3}(b). When the field is along $L_x$ ($\theta=0$), $h_{c0}$ is nearly constant, except for a slight decrease for small $L_y$ corresponding to the overall size suppression. The $\theta=0$ and $\theta=\pm\pi/2$ graphs intersect for $L_x=L_y$. Upon increasing $L_y>L_x$, the $\theta=\pm\pi/2$ direction becomes favorable. In this regime, the difference between the angles is less severe, since the size suppression is small.

In narrow superconducting strips with $L_x\gg L_y$, the system becomes effectively one-dimensional.
The limit where $L_x\to\infty$ and $L_y$ remains finite is shown in Fig.~\ref{fig:phase3}(c).
When the field is along the infinite direction, $h_{c0}$ equals the infinite film limit, which confirms that any suppression of $h_{c0}$ (compared to the infinite film) is a result of finite size.
This implies that, experimentally, the effect of SOC can be turned on and off in a narrow strip by rotating the in-plane field.
In the same limit, we compare the quasiclassical model presented here to an existing Ginzburg-Landau model \cite{Olde}. The angular dependency of the phase diagram close to $T_\text{c}$ can be recovered from thermodynamic arguments \cite{supp}, supporting the results presented here.
Since our calculation is in the diffusive limit (i.e. mean free path $\lambda \ll \xi$), we expects are results to be valid when $L,L_x,L_y \gtrsim \xi$.

We have shown that the paramagnetic limit $h_p$ of a thin-film superconductor is surpassed using Rashba SOC. We have demonstrated tunable superconductivity controlled by three parameters: the SOC strength, the sample geometry and the applied field direction. 
In shape anisotropic samples, the critical field is changed by rotating the field. The effect is visible over the whole temperature range up to $T_\text{c}$.

A possible experimental setup is a singlet superconducting thin-film with a thin heavy metal layer on top (bilayer thickness within $\xi$), such as Nb/Pt. In this setup, the Pt thickness determines the SOC strength. This can be extended to heterostructures with ferromagnets, such as Nb/Pt/Co, in which $h_{c2}$ is controlled by the ferromagnet exchange field.
Alternatively, to control the SOC within a single sample, the superconductor can be coupled to a two-dimensional chalcogenide in which the SOC is tuned by gating \cite{Premasiri,Afzal}.

\begin{acknowledgements}
This work is supported by the EPSRC through the Core-to-Core International Network ``Oxide Superspin''  (EP/P026311/1), the ``Superconducting Spintronics'' Programme Grant (EP/N017242/1) and the Doctoral Training Partnership Grant (EP/N509620/1).
\end{acknowledgements}

\bibliography{ref}

\begin{thebibliography}{41}
\bibitem [{not()}]{note_SOC}%
  \BibitemOpen
  \href@noop {} {}\bibinfo {note} {The SOC in our model is weak
  ($\sim\xi^{-1}$). Therefore, we do not take the modulated order parameter
  into account, which is valid for strong SOC
  \cite{Dimitrova,Houzet,Gardner}.}\BibitemShut {Stop}%
\end{thebibliography}

\end{document}